\documentclass{PoS}

\usepackage{bm}
\usepackage{graphicx}

\title{Kenneth Geddes Wilson}

\ShortTitle{Kenneth G. Wilson (1936--2013)}

\author{Andreas S. Kronfeld \\
    Theoretical Physics Department, Fermi National Accelerator Laboratory,\thanks{Operated by 
    Fermi Research Alliance, LLC, under Contract No.~DE-AC02-07CH11359 with the US DOE.}\hfill
    Batavia, Illinois, USA \\
    E-mail: \email{ask@fnal.gov}}

\abstract{A look back at Kenneth Wilson's contributions to theoretical physics, with some reminiscences of 
the professor I encountered at Cornell during the 1980s.\thanks{Photos from the American Institute of 
Physics (left) and Cornell University (right).}
\vspace{5em}
\begin{center}
\includegraphics[clip=true,trim=0 1.4cm 0 0,height=7cm]{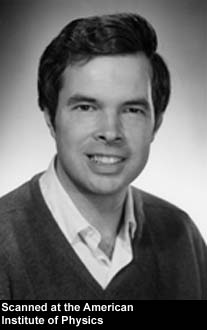}\quad
\includegraphics[height=7cm]{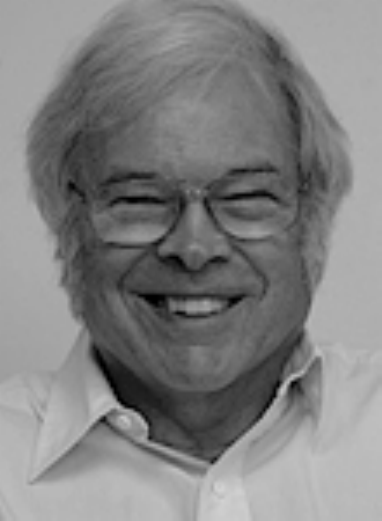}
\end{center}
}

\FullConference{31st International Symposium on Lattice Field Theory - LATTICE 2013 \\
		July 29 - August 3, 2013\\
		Mainz, Germany}

\begin{document}

\section{Prologue}

As part of the introduction to the 2013 Kenneth G.\ Wilson Award for Excellence in Lattice Field Theory, the
organizers have asked me to share some thoughts about Ken Wilson himself, to commemorate his contributions
to physics.
As most of you know, Ken succumbed to lymphoma on June 15, 2013, in Saco, Maine \cite{Cornell:2013,OSU:2013}.
Ken was not my advisor, but he was on my academic committee during my entire time in graduate school at
Cornell University.
He took part in administering my three oral examinations, I took his course on critical phenomena and the
renormalization group, and we collaborated with Peter Lepage and others on algorithms for lattice gauge
theory.

\section{Early Life}

Ken was born on June 8, 1936, to E.~Bright and Emily (n\'ee Buckingham) Wilson in Waltham, Massachusetts.
Wilson's father was an eminent chemist and co-author with Linus Pauling of a book on quantum
mechanics~\cite{Pauling:1935}.
Ken took to mathematics at a young age, computing cube roots in his head and learning calculus on his own.
He completed the third and fourth grades in one year, skipped eleventh grade, and graduated from the George
School, a Quaker boarding school in eastern Pennsylvania, at age sixteen \cite{Nobelprize.org}.
I grew up in this part of the world, attending George's archrival, Westtown, and always found it unusual
that George would let anyone skip a year.
En route from a hotel to Fermilab in 2004, I asked Ken about it, and he explained that he was bored, the
teachers could see it, and they decided to send him off to college as quickly as they could.

Ken went to Harvard, where he majored in mathematics, even though he had decided as a teenager to be a 
physicist. 
In 1956, he went to CalTech to obtain his Ph.~D.\ from Murray Gell-Mann.
Gell-Mann was often away, and Ken often said that he hiked and talked physics with Jon Matthews.
While still a graduate student, he went back to Harvard and the Society of Fellows (1959--1962), finishing 
the dissertation in 1961.
He went to CERN, and in 1963 Cornell offered Ken a faculty position unsolicited.
He liked the idea of being in the country and at a university with both a good physics department and a good 
folk-dancing club.

He received tenure in 1965 on the basis of the first (flawed---according to Wilson---and hence unpublished)
paper on the operator-product expansion~\cite{Wilson:1965zz} and a paper (submitted June 1965) on
renormalization~\cite{Wilson:1965xy}.
At Cornell in the 1980s, a story circulated that Gell-Mann wrote a letter testifying to Ken's brilliance, but
cautioning that tenure was premature.
The second half of the story is that Hans Bethe told his colleagues to read only the first paragraph.
% Ken's father, Bright Wilson, supposedly lamented (with irony) that the published paper was a pity, because it had
% prevented Ken from showing it possible to receive tenure without publishing.

\section{Science and Language}

After five years of work, analyzing the OPE idea~\cite{Wilson:1968rg} in the solvable Thirring
model~\cite{Wilson:1970tm} and in perturbation theory~\cite{Wilson:1970pt}, Ken starting publishing landmark
papers using the OPE to link renormalization to (broken) scale invariance.
In the early 1970s, his output became a torrent, in sections D~\cite{Wilson:1971rg,Wilson:1971tv} and
B~\cite{Wilson:1971bg,Wilson:1971dh} of the Physical Review.
His papers always generously give credit, tracing, for example, the idea of broken scale invariance in
the strong interactions to Gerhard Mack~\cite{Mack:1968zz} and Hans Kastrup~\cite{Kastrup:1966sc}.
Ken's influence on language also began: he introduced the term ``anomalous dimension'' for the part of the
scaling dimension arising from quantum-mechanical effects.
Others started calling the $c$-number coefficient functions of the OPE ``Wilson coefficients.''

The two 1971 back-to-back papers in Phys.~Rev.~B~\cite{Wilson:1971bg,Wilson:1971dh} began Ken's assault on
critical phenomena.
Here again, his way of talking about physics led to commonplace terms
(here and below \emph{italics} are added to emphasize terms coined by Wilson):
\begin{quote}
    Equations (9) and (10) are the renormalization-group equations suggested by the Kadanoff 
    \emph{block} [\emph{spin}] picture~\cite{Wilson:1971bg}
\end{quote}
even though Leo Kadanoff had called the subdivisions of his lattice ``cells'' not ``blocks.''
Further:
\begin{quote}
    To derive the recursion formula, the variables $\sigma_L(\bm{k})$ for $0.5L\le|\bm{k}|\le L$
    will be \emph{integrated out}~\cite{Wilson:1971dh}
\end{quote}
and, indeed, in order to ``integrate out'' many integrals are carried out in these pages.

Wilson's work in the 1970s represents an amazing period of cross-fertilization.
With Michael Fisher, he introduced an expansion in spatial dimension $d$ around $4$ to compute critical
exponents~\cite{Wilson:1971dc}.
The title, ``Critical Exponents in 3.99 Dimensions,'' which Fisher attributes to Ken, shows a man who is not
afraid of floating-point numbers.
Soon afterwards Ken introduced Feynman diagrams for the same purpose~\cite{Wilson:1971vs}, collaborating with
young field theorists \'Edouard Br\'ezin and David Wallace~\cite{Brezin:1972fc,Brezin:1972fb}.
Their collaboration grew out of a series of lectures delivered at Princeton University.
John Kogut took notes, leading to a monumental review article~\cite{Wilson:1973jj} that everyone should read,
even now.
In the midst of this flurry of activity in condensed matter physics, Ken still thought about particle
physics, for example putting the OPE on a firmer mathematical footing with Wolfhart
Zimmermann~\cite{Wilson:1972ee}.

And so, when asymptotic freedom was discovered~\cite{Gross:1973id,Politzer:1973fx}, Ken refocused his
attention on quantum field theory for the strong interactions.
At Lattice 2004, Ken explained that he was eager to carry out research on QCD, but that he hadn't learned the
intricacies of gauge-fixing and so on~\cite{Wilson:2004de}.
It seemed easier to formulate a lattice field theory with exact gauge symmetry, which his foray into critical
phenomena had prepared him to do.
The resulting paper is the watershed marking the start of lattice gauge theory~\cite{Wilson:1974sk}: it gave
a simple picture of confinement, based on a theoretical tool now known throughout the study of gauge field
theory as the ``Wilson loop.''

Some might argue that ``Confinement of Quarks''~\cite{Wilson:1974sk} was his second-most influential paper of
1974.
His solution of the Kondo problem~\cite{Wilson:1974mb} had a big impact, not least on language:
\begin{quote}
    If a particular parameter in the initial Hamiltonian causes amplification (i.e., $\lambda_i$
    is greater than 1), it is called a ``\emph{relevant variable}.''
    A~parameter whose effect is deamplified is called ``\emph{irrelevant}.'' \ldots\ 
    A \emph{marginal variable} is one which is neither amplified nor deamplified by a renormalization group 
    transformation. \ldots\ 
    [T]he field theoretic methods of Gell-Mann and Low and Callan and Symanzik are based entirely on the 
    special behavior of \emph{marginal variables}. 
\end{quote}
Eight years later, upon receiving the Nobel Prize in physics, Ken referred to the Kondo problem as an important
application of the renormalization group beyond critical phenomena~\cite{Wilson:1993dy}.

The Kondo paper is full of ``integrating out'' to derive effective Hamiltonians at longer length scales.
In particle physics, the Wilsonian renormalization group again leads to a concept of effective field theory:
a (perhaps slightly) different quantum field theory at every scale.
Around the same time, Steven Weinberg (certainly not an obscure figure) introduced a related concept, based
on unitarity, analyticity, and cluster decomposition~\cite{Weinberg:1978kz}.
In Wilson's concept, symmetries lead to universal behavior at long distances; in Weinberg's, symmetries are
used to impose or derive constraints on a posited long-distance effective Lagrangian.
With Weinberg, one can skip the tedium of performing any integrals.
Even so, authors often draw on Weinberg's line of reasoning yet use Wilson's vocabulary and talk about
``integrating out degrees of freedom.''

\begin{figure}[b]
    \includegraphics[width=0.48\textwidth]{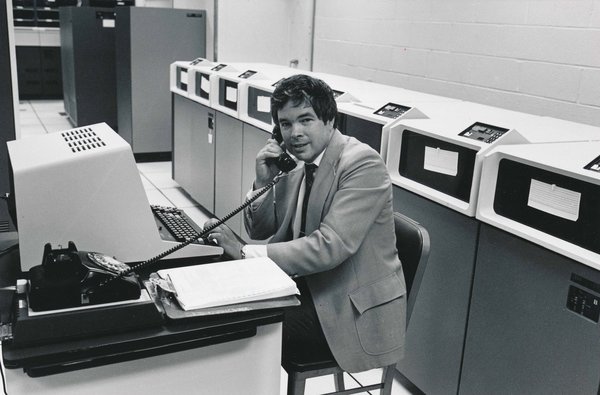}\hfill
    \includegraphics[width=0.48\textwidth]{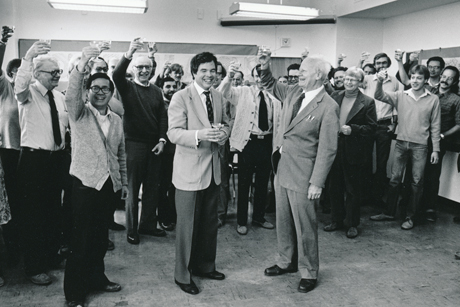}
    \caption{Left: Ken Wilson at a computer.
        Young readers may not know that the large cabinets house disk drives, the large box with a keyboard 
        is a VT100 terminal, and the box on the desk in the foreground is a old computer communications 
        device called a modem.  
        The funny-looking black thing on top of the modem (Ken is holding part of it, connected with a cord) 
        is a twentieth-century telephone.
        Right: Nobel celebration in Newman Laboratory, with Ken (center) and a beaming Hans Bethe 
        (center right).  
        Tung-Mow Yan is also in the foreground, left.
        To the far right, classmate Ray Renken, and, partly obscured by Ray's glass, Peter Lepage.
        The mass of curly hair on the left (in the back corner, obscured) is probably mine.
        Photographs from Cornell University.}
    \label{fig:fig}
\end{figure}

Since his undergraduate days, Ken liked computers and computing, and he often credited Jon Matthews with
teaching him to use the CalTech computer.
By 1980, he urged theorists to exploit the computer for understanding physics.
He met with resistance from some physicists.
Through his work with the Lax Panel on Large-Scale Computing in Science and Engineering~\cite{Lax:1982},
he nevertheless had an influence on policy makers.
In a ten-page supplemental chapter, Wilson outlined and justified recommendations that found their way into
the main report.
He wrote:
\begin{quote}
    The immediate national\footnote{The report was written under the sponsorship of the U.S.
    Department of Defence and the National Science Foundation, in cooperation with the Department of Energy 
    and the National Aeronautics and Space Administration.} needs are the following:
    \begin{enumerate}
        \item[1)] A national network linking all scientists involved in open basic research, vastly 
            generalizing the existing Arpanet and Plasma fusion energy networks.
        \item[2)] A development program in support of large scale scientific computing, encompassing hardware, 
            systems software, and algorithm development and carried out as a collaboration between 
            knowledgeable members of the scientific community and the computing industry.
        \item[3)] Building an adequate equipment base (computers, peripherals, and network access) for 
            training and theoretical research in universities.
        \item[4)] Providing adequate access by researchers on the network to special and general purpose 
            facilities at the National Laboratories and elsewhere, for computing needs that go beyond the 
            base level.
    \end{enumerate}
\end{quote}
We can recognize these technologies---international in scope, of course---and cannot imagine research 
without them.
Ken dove into computer evangelism with all his energy, for example posing for an amusing photograph, 
reproduced here on the left of Fig.~\ref{fig:fig}.

\section{Reminiscences}

At Cornell, new graduate students found that Ken's reputation preceded him.
The other professors were always telling us how great he was.
Being accustomed to ``great professors'' among my father's colleagues, I~was a bit surprised to find him so
kind and friendly, in an introverted way, and even modest.
But not false modesty: everyone knew he was the smartest guy in the room, but he didn't need to prove it and
didn't want to dwell on it.
His sense of humor consisted of making statements that sounded oracular, unless you understood quantum field
theory, in which case the statement was a kind of pun, but one that illuminated physics.
When Ken found something funny, he had a wonderful, impish smile.

In Ithaca, the Nobel Prize did not come as a surprise.
Fisher, Kadanoff, and Wilson had shared the 1980 Wolf Prize, and many just assumed that they would receive 
the Nobel in due course.
The surprise, then, was that the Nobel committee saw in Wilson's work a broader revolution in scientific 
thought.
The citation is for ``for his theory for critical phenomena in connection with phase transitions,'' but the 
press release notes a wider range of application.
At Cornell, the particle physicists had a party in Newman Laboratory (see Fig.~\ref{fig:fig}, right), and a 
bit later the physics department had another party in Clark Hall.
Hans Bethe was beaming like a proud father.
Michael Fisher graciously explained why the Nobel committee saw the physics correctly.

At oral examinations, Ken did not ask questions.
Instead he would direct students to ``tell me about hydrogen,'' or, with Ray Renken, ``tell me about
molecules.'' %
Ray and Ken ended up in a dialogue about diatomic molecules and potentials that could describe their
vibrational modes.
Even though his technique was well-known among us graduate students, it could still be flabbergasting.
Taking my A~Exam (the exam at Cornell marking the transition from coursework to research) soon after Ray and
a few others, I studied until I knew every simple problem by heart.
Ken chose the simplest of the simple: ``Tell me about the harmonic oscillator!'' %
I leapt to the blackboard and started to solve it with the ladder operators.
A pained expression came over Ken's face; he was not interested in that.
We ended up having a nice chat about a simple way to infer the equal level spacing from the symmetry of
the Hamiltonian under the interchange of $p$ and~$x$.

We students did not consider Ken to be a good classroom lecturer.
Part of the problem was the impedance mismatch between his mind and ours.
In the case of his Autumn 1982 course on critical phenomena, he was also somewhat distracted.
By then I had learned not to filter his words through my own mind, but to write them down as accurately as I
could.
(In seminars and informal situations, I memorized his sentences.) %
The notes turned out fine.
When Yuko Okamoto (who worked on $g-2$ with Toichiro Kinoshita, later on lattice gauge theory, and now on
protein folding) visited Fermilab in the early 1990s, he asked to see the notes.
We found them lucid and easy to understand.
Our minds had matured, so the impedance mismatch had lessened.

Contrary to the folklore stemming from his remarks at a panel discussion at Lattice 1986, Ken remained
interested in lattice gauge theory after shifting his research to other topics.
I think Ken's remarks in 1986 and 1989~\cite{Wilson:1989ax} have been overinterpreted and misinterpreted.
While it is disappointing that he decoupled from the lattice community, he did not lose interest with the 
work.
He kept in contact with a few people, especially Peter Lepage.
The key message in Ken's remarks is to pursue---as he did---a broad range of subjects in your research.
% Is that a surprise, when you look at his most fruitful period? %
% Isn't our own research more fun when colleagues---experimenters or nonlattice theorists---like our results?

Indeed, Ken's Lattice 2004 talk~\cite{Wilson:2004de} contains a salute to the progress that had been made,
especially in the earliest unquenched calculations.
When \emph{Science} published the BMW Collaboration's paper on the hadron mass spectrum in
2008~\cite{Durr:2008zz}, Ken called me to discuss it.
He right away asked me to convey his congratulations to the authors, whom he didn't know.
The next question was whether or when lattice QCD would be precise enough to detect new forces.
I mentioned the tension in $f_{D_s}$~\cite{Dobrescu:2008er}, but he was more interested in deviations in the
hadron masses themselves.
He then went on to raise issues in particle physics and cosmology that had been bothering him.
Not wanting to filter, I wrote down as much as I could, considering that I had to reply and respond while
taking notes.
Ken's interests seemed as much philosophical as scientific.
He still read papers but, in his retirement in Maine, was not in touch with the unwritten side of science.
I clearly remember him reacting against a subject that, in his mind, was locked into a single
implausible-sounding paradigm, noting that it wouldn't be the first time scientists were completely misguided.
Unfortunately, I've misplaced my notes; although I believe I remember what the subject was, it seems better
not to reveal it without looking at my notes.

\section{Epilogue}

What happened after Ken Wilson's revolutionary spike in creativity and productivity? For a while, important
work on lattice gauge theory appeared in the proceedings of summer schools, for example Wilson
fermions~\cite{Wilson:1975id} and early Monte Carlo calculations~\cite{Wilson:1977nj}.
Much of this work concerns renormalization beyond perturbation theory and the marriage of numerical
simulation with effective field theories to obtain physical results.
At Cornell in 1983, Ken and Peter recruited a group of students and postdocs to investigate of numerical
algorithms for lattice gauge theory~\cite{Batrouni:1985jn,Davies:1987vs}.
In 1988, Ken moved to the Ohio State University, where he split his research time between light-front
QCD~\cite{Wilson:1994fk} and social issues such as science education.
At OSU, he wrote twelve physics publications with average citation count 94~\cite{inSPIRE:hep}.
His study of physics education was also fruitful, leading to a book~\cite{Wilson:1994re} and several research
papers, for example with Constance Barsky~\cite{Barsky:1998rd}.

My way of making sense of Ken Wilson's career is to view him as a philosopher as well as a physicist.
He thought deeply, not only about physics but also about how to think deeply.
It is perhaps for this reason that his work, beyond its excellence as theoretical physics, influenced our way of talking
and thinking so much.
The Lattice 2004 writeup has an explicitly philosophical tone; turn to any interview with Wilson in his
latter years, such as the one in \emph{Candid Science~IV} \cite{Hargittai:2004iv}, and you will find deeper 
and even more fascinating thoughts on the meaning of science and its place in society.

\acknowledgments
Cornell University organized a memorial symposium on November 16, 2013, which 
brought back some memories that I've woven into the narrative presented at the conference.

\appendix
\section{The Kenneth G. Wilson Award for Excellence in Lattice Field Theory}

The International Advisory Committee (IAC) of the 2011 International Symposium on Lattice Field Theory
introduced the Ken Wilson Lattice Award.
For 2013, some changes in process, target, and award name have been introduced.
In particular, the new version is nomination-driven, selected by an appointed committee, and awarded to a
promising young scientist.

The 2013 award, \emph{for significant contributions to our understanding of baryons using lattice QCD and
effective field theory}, is awarded to Dr.\ Andr\'e Walker-Loud (now assistant professor at the College of
William and Mary).
Prof.\ Walker-Loud has given an account of his work in these proceedings~\cite{Walker-Loud:2013}.

\section{2013 Announcement}
\noindent
From the \href{http://www.lattice2013.uni-mainz.de/219_ENG_HTML.php}{Lattice 2013 web site}:
\begin{quote}
The prize is given annually and will consist of a certificate citing the contributions of the recipient, a
modest monetary award, and an invitation to present the cited work in a short plenary talk at the
International Symposium on Lattice Field Theory.
The registration fee for the conference will be waived.

This award recognizes outstanding physicists who are within seven years of Ph.~D.\ at the time of nomination,
plus any career breaks.
The research recognized could either be a single piece of work, or the sum of contributions.

Nominations should consist of a nominating letter, a seconding letter, and up to two supporting letters; the
nominee's CV; and a list of relevant publications (up to ten).
The nominating letter should include a proposed citation (one sentence) justifying the award.
All letters should explain and justify why the nominee should receive the award.
Nominations will be considered for three years (or until receiving the award).

Self-nominations will not be considered.
Otherwise, anyone may submit a nomination.
Recipients are not eligible for a second award.
Nominations of scientists from all nations are solicited.

For nominations submitted for the 2013 award, nominees must have received a Ph.~D.\ (or equivalent) after
December 31, 2005 (with suitable adjustment in cases with career breaks).
Send nominations to the Selection Committee chair by April 5, 2013.
\end{quote}
The 2013 Selection Committee consisted of Andreas Kronfeld (Chair), Shoji Hashimoto, Karl Jansen (2011
recipient), Stephen Sharpe (Vice-Chair), and Maria Paola Lombardo.
Prof.\ Sharpe will chair the 2014 committee, and Dr.\ Lombardo will serve again.
Prof.\ Norman Christ, a 2012 recipient, will join the committee, as will two members appointed by the IAC.

The 2013 IAC, award founder Dr.\ Pavlos Vranas, and the lattice-field-theory community hope that the Kenneth
G.\ Wilson Award for Excellence in Lattice Field Theory will commemorate the man and his contributions to
theoretical physics.

\end{document}